\input{aipcheck}
\documentclass[
    ,final            
  ]
  {aipproc}

\layoutstyle{6x9}

\begin{document}

\title{Relativistic Quark-Model Results \\
       for Baryon Ground and Resonant States\footnote{Copyright (2008) American Institute
       of Physics. This article may be
  downloaded for personal use only. Any other use requires prior permission of
  the author and the American Institute of Physics. The article
  appeared in AIP Conf. Proc. Vol. 1056, pp. 15-22 (2008) and may be found at
  http://link.aip.org/link/?APCPCS/1056/15/1.}}      


\author{W. Plessas and T. Melde}{
  address={Theoretische Physik, Institut f\"ur Physik, Karl-Franzens-Universit\"at Graz, \\
Universit\"atsplatz 5, A-8010 Graz, Austria}
}

\begin{abstract}
Latest results from a study of baryon ground and resonant states within relativistic constituent quark models are reported. After recalling some typical spectral properties,
the description of ground states, especially with regard to the nucleon and hyperon
electromagnetic structures, is addressed. In the following, recent covariant predictions
for pion, eta, and kaon partial decay widths of light and strange baryon resonances below
2 GeV are summarized. These results exhibit a characteristic pattern that is distinct from nonrelativistic or relativized decay studies performed so far. Together with a detailed analysis of the spin, flavor, and spatial structures of the wave functions, it supports a new and extended classification scheme of baryon ground and resonant states
into SU(3) flavor multiplets. 
\end{abstract}

\maketitle

\section{RELATIVISTIC CONSTITUENT-QUARK MODELS}

For the results discussed here we employ two kinds of relativistic constituent-quark
models (RCQMs). These are the Goldstone-boson-exchange (GBE) RCQM~\cite{Glozman:1998ag} and
a variant of the one-gluon-exchange (OGE) RCQM presented in ref.~\cite{Theussl:2000sj}.
They are defined through a relativistically invariant mass operator $\hat M$ that is
treated in the framework of Poincar\'e-invariant quantum mechanics. The solution of its
eigenvalue equation
\begin{equation}
    {\hat M}\left|V,M,J,\Sigma\right>
    =M\left|V,M,J,\Sigma\right>
    \label{eq:mass}
\end{equation}
leads to the mass $M$ and eigenstate $\left|V,M,J,\Sigma\right>$ of a baryon ground or
resonant state, characterized by intrinsic spin $J$ with $z$-component $\Sigma$; the
$\left|V,M,J,\Sigma\right>$ are simultaneously eigenstates of the velocity operator
$\hat V^{\mu}$ and also of the momentum operator $\hat P^{\mu}=\hat M\hat V^{\mu}$.

For our calculations of baryon reactions, in particular of the elastic electromagnetic
and axial form factors of the nucleons and the mesonic decays of the baryon resonances
considered below, we adhere to the point form of Poincar\'e-invariant
quantum mechanics, since it allows to produce manifestly covariant results. This is
essentially a consequence of the generators of Lorentz transformations to be independent
of interactions; the only interaction-dependent generators of the Poincar\'e group are the
four components of the momentum operator $\hat P^{\mu}$.

The mass-operator eigenvalue equation~(\ref{eq:mass}) is solved with the stochastic
variational method (SVM)~\cite{Suzuki:1998bn}, which allows to achieve very accurate mass eigenvalues and also the most general dependence of baryon wave functions on color, flavor,
spin, and spatial variables. More details on the formalism can be found, e.g., in
ref.~\cite{Melde:2008yr} and references therein. 

\section{BARYON SPECTRA}

\begin{figure}[h]
\includegraphics
[height=8cm]{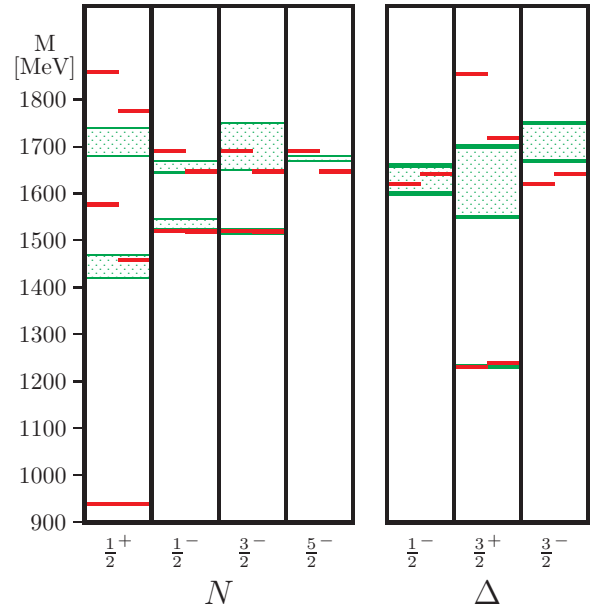}
\caption{Energy levels (red solid lines) of the lowest $N$ 
and $\Delta$ states with total angular momentum and parity $J^P$ for the OGE
(left levels) and GBE (right levels) RCQMs in comparison to
experimental values with uncertainties~\cite{PDBook}, represented as (green)
shadowed boxes.}
\label{fig:NDel}
\end{figure}

\begin{figure}
\includegraphics
[height=8cm]{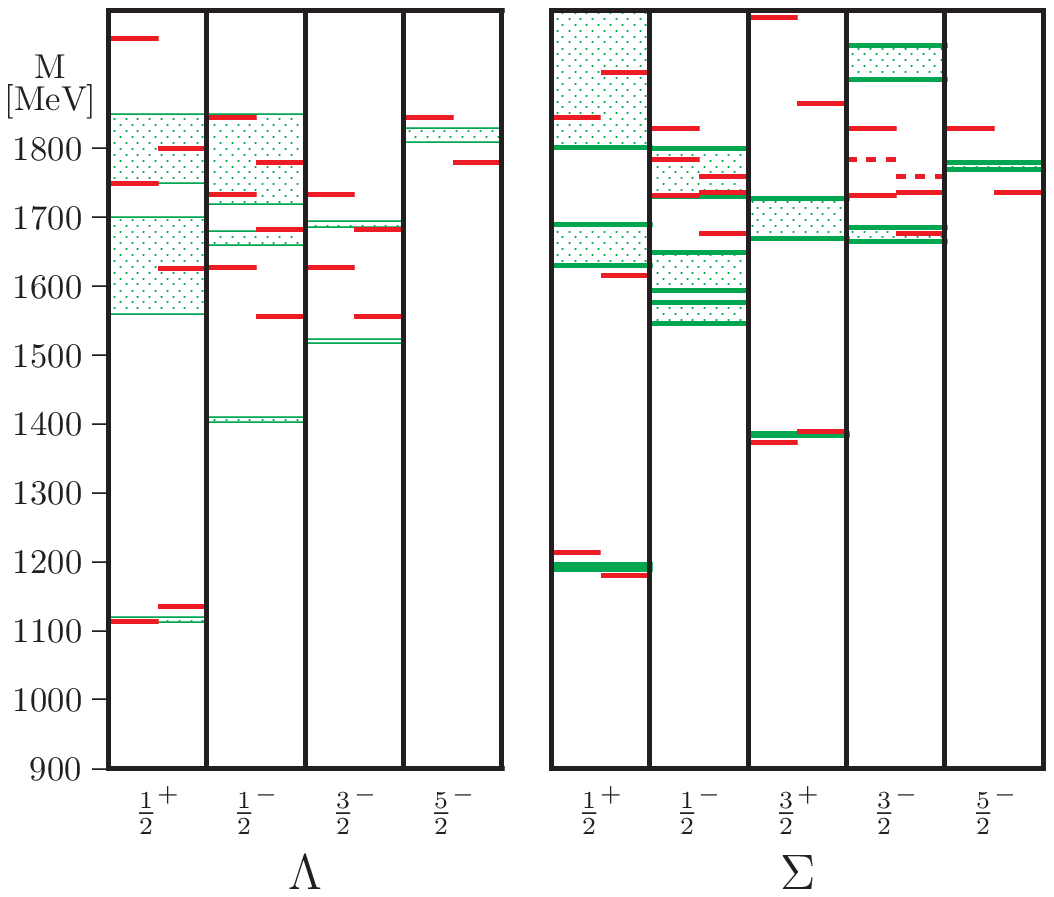}
\hspace{0.05cm}
\includegraphics
[height=8cm]{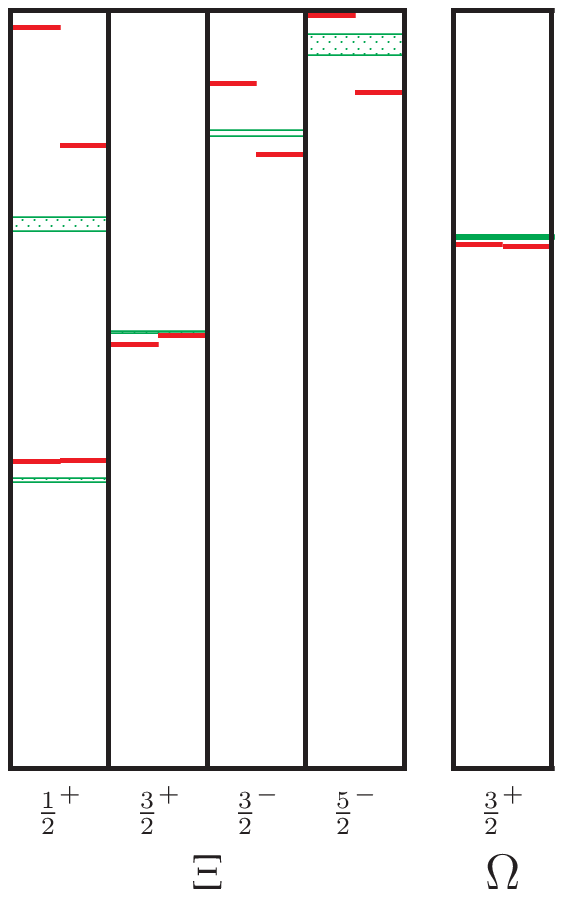}
\caption{Sames as in fig.~\ref{fig:NDel} for the lowest $\Lambda$, $\Sigma$, $\Xi$, 
and $\Omega$ states. The dashed lines in the $J^P=\frac{3}{2}^-$ $\Sigma$ spectrum
represent (decuplet) eigenstates, for which there is no experimental counterpart yet.}
\label{fig:LaSig}
\end{figure}

The invariant mass spectra of the two types of RCQMs are shown in figs.~\ref{fig:NDel}
and~\ref{fig:LaSig} for the baryon states below $\approx 2$ GeV considered
here\footnote{The complete spectra of the RCQMs can be found in
refs.~\cite{Glozman:1998ag,Theussl:2000sj,Glozman:1998fs}.}.
Some striking features of the spectra are immediately evident. Here, we only hint to
the wrong level ordering of the OGE
RCQM with regard to the $\frac{1}{2}^+$ Roper resonance $N(1440)$ and the $\frac{1}{2}^-$
resonance $N(1535)$ in the $N$ spectrum (fig.~\ref{fig:NDel}) and the failure of both RCQMs
in describing the first excitation $\Lambda(1405)$ in the $\Lambda$ spectrum
(fig.~\ref{fig:LaSig}).


\section{ELECTROWEAK STRUCTURE OF BARYON GROUND STATES}

\subsection{Electromagnetic nucleon form factors}

A first test of the mass-operator eigenstates $\left|V,M,J,\Sigma\right>$ concerns the
elastic electromagnetic from factors of the nucleon. In this context a simplified
current operator according to the point-form spectator model (PFSM) has been used so far.
It means that the virtual photon couples only to a single constituent quark in the nucleon.
Nevertheless the PFSM current represents an effective many-body operator~\cite{Melde:2007zz}.

\begin{figure}
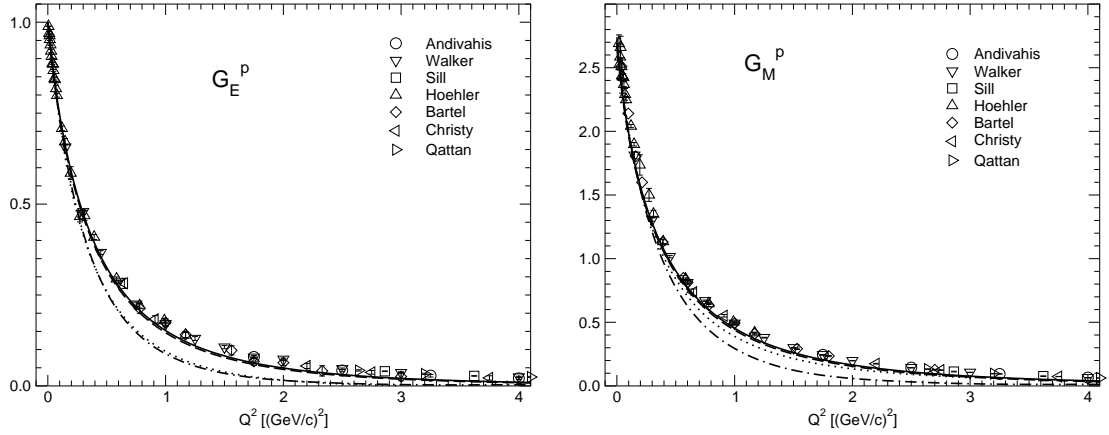

\includegraphics[height=5.7cm,clip=]{gep_BCN.eps}
\hspace{0.3cm}
\includegraphics[height=5.7cm,clip=]{gmp_BCN.eps}
\caption{\label{fig:proton_comp}
Electric and magnetic form factors of the proton as predicted by the
GBE (full line) and OGE (dashed line) RCQMs along the PFSM approach; in addition
the results for the case with only the confinement potential (inherent in the
GBE RCQM) are given (dash-dotted line). For comparison also the predictions of
the II RCQM (dotted line) after ref.~\cite{Merten:2002nz} are shown.
Experimental data are from
refs.~\protect\cite{Lung:1993bu,Markowitz:1993hx,%
Rock:1982gf,Bruins:1995ns,Gao:1994ud,Anklin:1994ae,Anklin:1998ae%
,Xu:2000xw,Kubon:2001rj,Xu:2002xc} 
and~\protect\cite{Andivahis:1994rq,Walker:1989af,Sill:1993qw,Hohler:1976ax,Bartel:1973rf,%
Christy:2004rc,Qattan:2004ht}.
}
\end{figure}

\begin{figure}
\includegraphics[height=5.7cm,clip=]{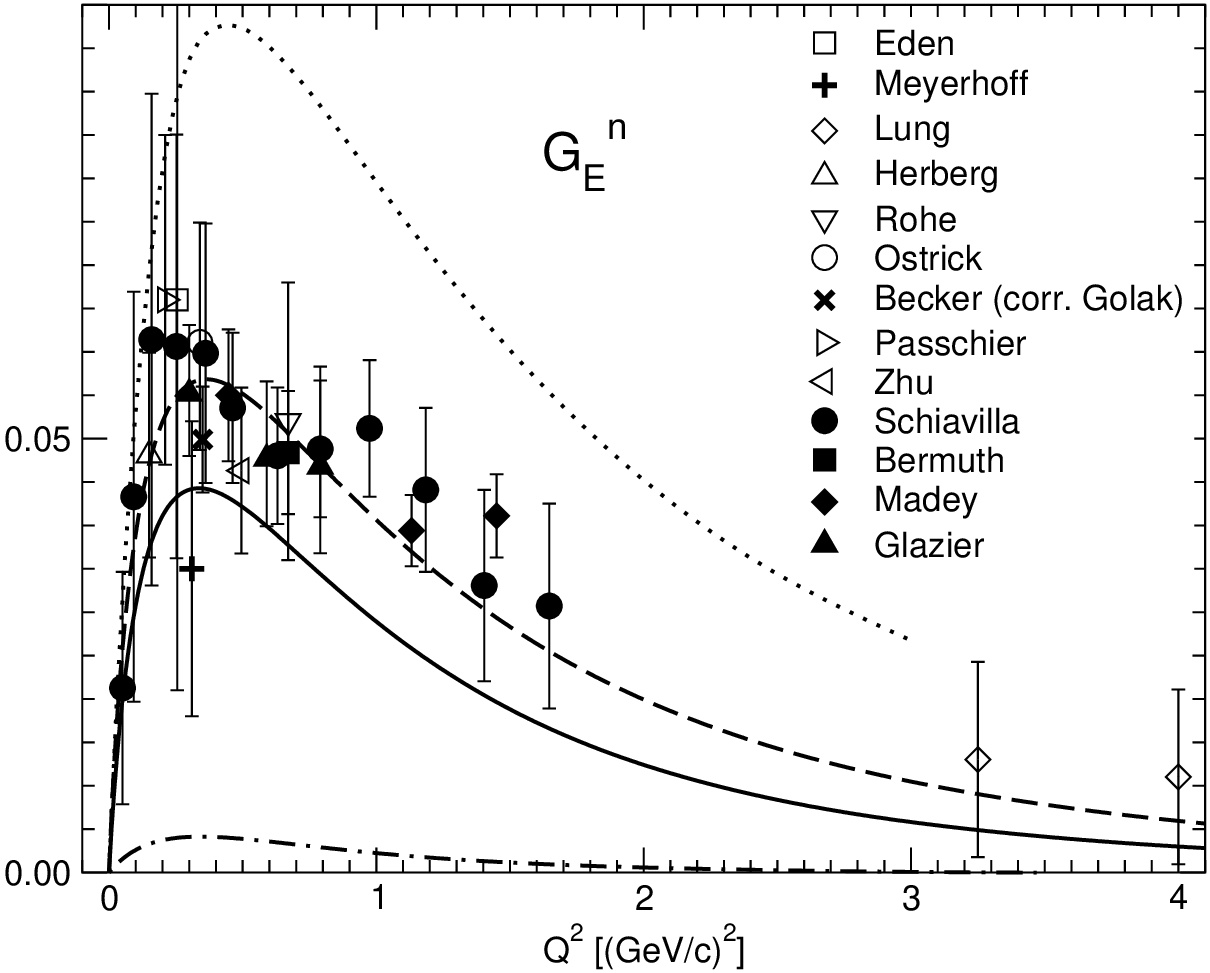}
\hspace{0.3cm}
\includegraphics[height=5.7cm,clip=]{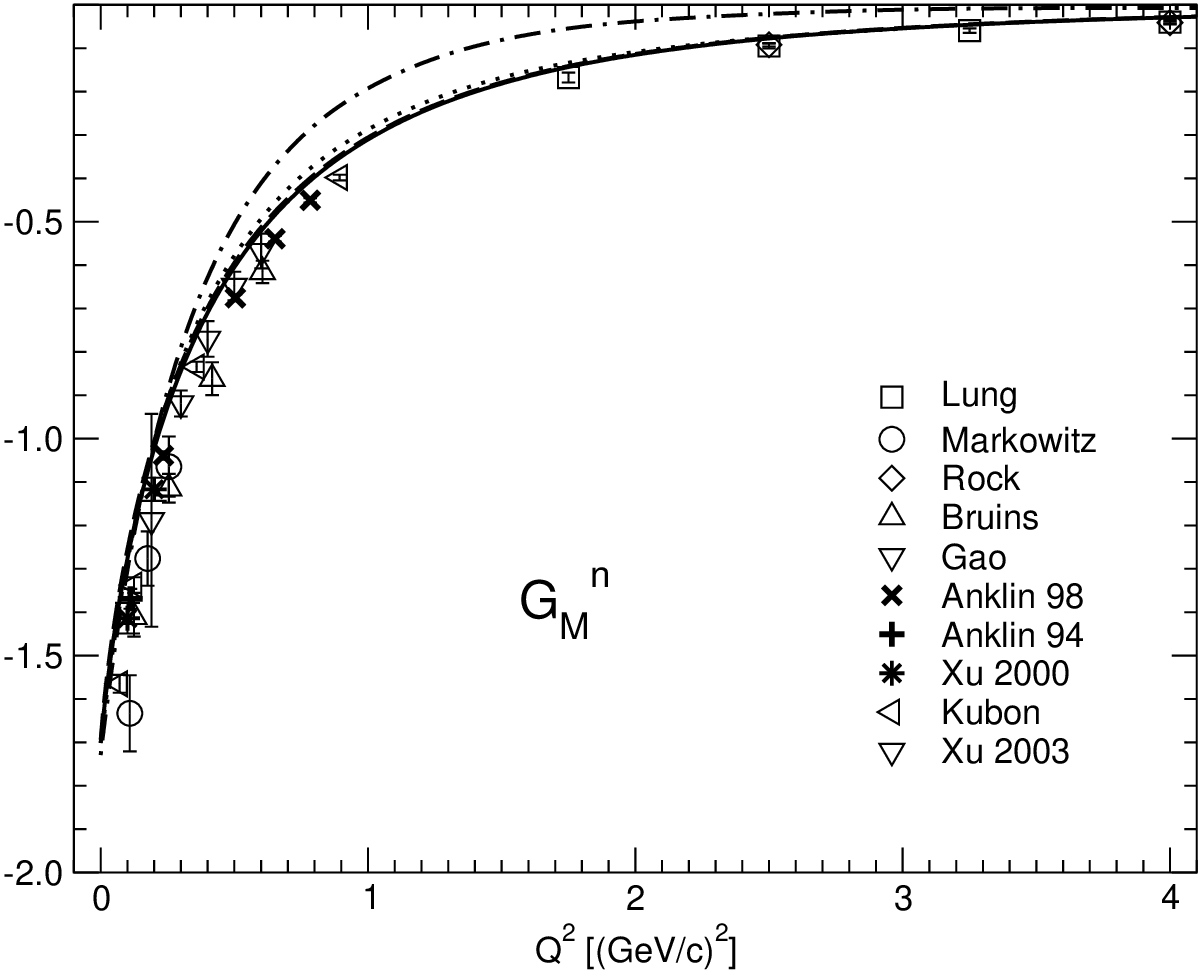}
\caption{\label{fig:neutron_comp}
Same as in fig.~\ref{fig:proton_comp} but for the neutron. Experimental data are from
refs.~\protect\cite{Lung:1993bu,Markowitz:1993hx,%
Rock:1982gf,Bruins:1995ns,Gao:1994ud,Anklin:1994ae,Anklin:1998ae%
,Xu:2000xw,Kubon:2001rj,Xu:2002xc,%
Eden:1994ji,Meyerhoff:1994ev,Herberg:1999ud,Rohe:1999sh,%
Ostrick:1999xa,Becker:1999tw,Passchier:1999cj,Zhu:2001md,%
Schiavilla:2001qe,Bermuth:2003qh,Madey:2003av,Glazier:2004ny}.
}
\end{figure}

The covariant predictions of the GBE and OGE RCQMs for the electric and magnetic form
factors of the proton and the neutron are shown in figs.~\ref{fig:proton_comp}
and~\ref{fig:neutron_comp}. For momentum transfers up to $\approx 4$ GeV$^2$, where
we may assume a quark-model description to be reasonable, the results of both the GBE
and OGE RCQMs are found in surprisingly good agreement with the available experimental
data. A nonrelativistic calculation along the usual impulse approximation fails
completely~\cite{Wagenbrunn:2000es,Boffi:2001zb}. Similarly a relativistic
calculation performed along an instant-form spectator model, constructed in anology to
the PFSM, is found to be far off the experimental
data~\cite{Melde:2007zz}. While the detailed behaviour of the nucleon wave function is
of lesser influence, the correct treatment of relativistic effects appears to be most
essential. This observation is further supported by the comparison with the relativistic
results for the nucleon form factors obtained by the Bonn group with their instanton-induced
(II) RCQM~\cite{Loring:2001kx} in a completely different
Bethe-Salpeter approach~\cite{Merten:2002nz} (also shown in figs.~\ref{fig:proton_comp}
and~\ref{fig:neutron_comp}). On the other hand, a simplistic nucleon wave function, such
as the one obtained with a confinement potential only, is also not adequate. Since it misses
important mixed-symmetric spatial components, it yields an almost zero result in particular
for the neutron electric form factor (see fig.~\ref{fig:neutron_comp}).


A corresponding behaviour is found for the nucleon electromagnetic radii and magnetic
moments: The direct PFSM predictions of the RCQMs are very close to the experimental
data both for the proton and the neutron~\cite{Melde:2007zz,Berger:2004yi}. Even for these
observables, which relate to momentum transfers $Q^2 \rightarrow 0$, relativistic effects
are high importance, and a nonrelativistic theory is by no means adequate.

Quite similar results are obtained for electric radii and magnetic moments also of the
other baryon ground states, where a comparison to experiment
is possible~\cite{Berger:2004yi}.

\subsection{Axial nucleon form factors}

With regard to the nucleon axial and induced pseudoscalar form factors, $G_A$ and $G_P$,
the situation is quite similar to the elastic electromagnetic form
factors~\cite{Boffi:2001zb,Glozman:2001zc}. Only the covariant predictions of the RCQMs
come close to the experimental data. Again the PFSM results and the ones obtained
in the Bethe-Salpeter approach are alike~\cite{Plessas:2004fn}. The nonrelativistic
impulse approximation is not acceptable. In case of the induced pseudoscalar form factor
$G_P$ the inclusion of the pion pole term is crucial. It can naturally be implemented
for the GBE RCQM, in line with the dynamics in its hyperfine interaction (pseudoscalar
boson exchange).
\begin{figure}
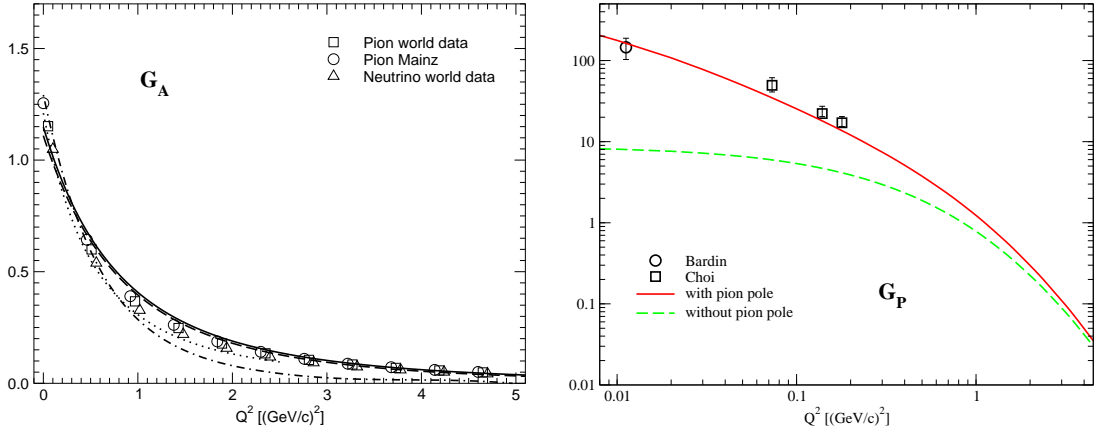

\includegraphics[height=5.7cm,clip=]{ga_cqms.eps}
\hspace{0.3cm}
\includegraphics[height=5.7cm,clip=]{gp_prd.eps}
\caption{Nucleon axial and induced pseudoscalar form factors $G_A$ and
$G_P$, respectively. For $G_A$ the denotation of the curves is the same as in
figs.~\ref{fig:proton_comp} and~\ref{fig:neutron_comp}. The experimental data 
are shown assuming a dipole parameterization with the axial 
mass value $M_A$ deduced from pion electroproduction (world average: 
squares, Mainz experiment~{\protect\cite{mainz}}: circles) and from neutrino
scattering~{\protect\cite{neut}} (triangles). For $G_P$, results from the GBE RCQM are
shown with and without pion-pole contribution. The corresponding experimental data are
from ref.~{\protect\cite{gp-exp}}.
\label{fig:axial}
}
\end{figure}

\section{Strong Decays}

\begin{figure}
\includegraphics[angle=0,clip=,height=10cm]{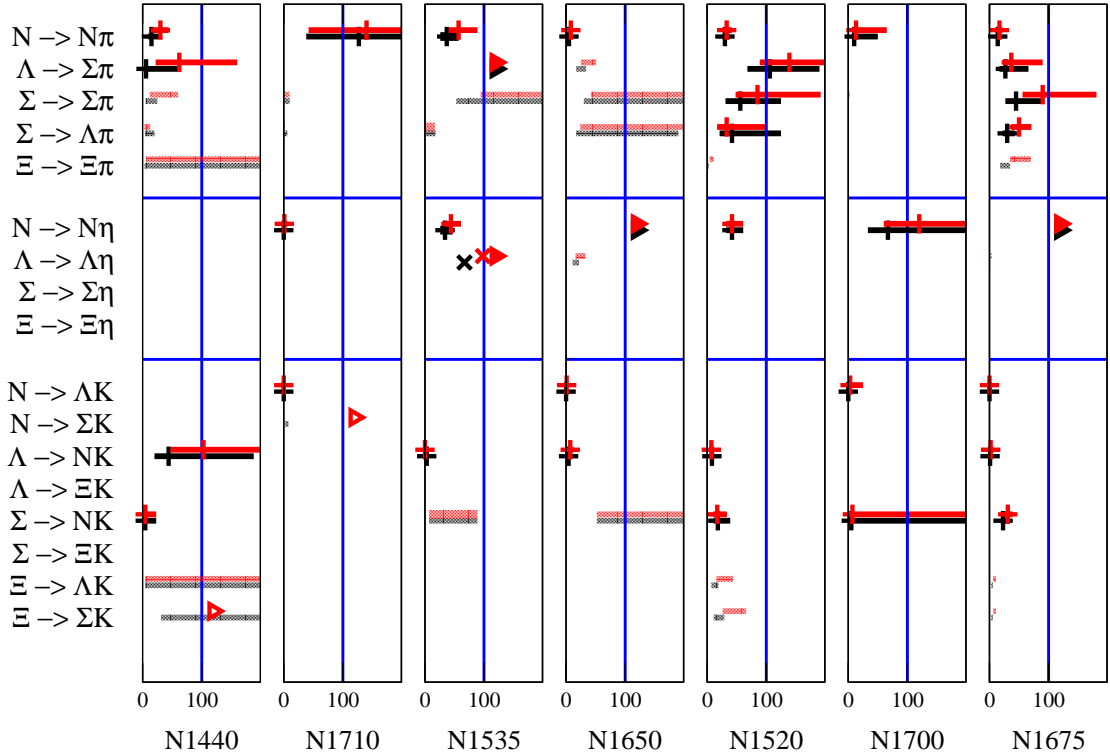}
\caption{Predictions for partial $\pi$, $\eta$, and $K$ decay widths of the GBE
(black/lower entries) and OGE (red/upper entries) RCQMs from the PFSM calculation for
certain flavor octets (cf. table~\ref{tab:multiplet_oct} in the next section).
The results shown by + crosses are presented as percentages of the best estimates
for experimental
data reported by the PDG~\cite{PDBook}, with the horizontal lines showing the
experimental uncertainties. In case of shaded lines without crosses the PDG gives only total
decay widths, and the theoretical results are represented relative to them. The
triangles point to results outside the plotted range. For the particular decay
$\Lambda(1670) \rightarrow \Lambda \eta$ in addition to the theoretical masses also
experimental ones were used, and the corresponding results are marked by $\times$ crosses.}
\label{fig:decay_graph_OGE_GBE_oct}
\end{figure}

Recently, the relativistic study of all single-meson decay modes of the light and
strange resonances below $\approx 2$ GeV has been completed in the approach adopting a
PFSM decay operator~\cite{Melde:2005hy,Melde:2006yw,Sengl:2007yq}. The covariant results
for the partial $\pi$, $\eta$, and $K$ decay widths show a typical behaviour in the sense
that they usually underestimate the experimental data. The situation is exemplified for
octet baryon resonances in fig.~\ref{fig:decay_graph_OGE_GBE_oct}. Similar patterns
are obtained for the decuplet and singlet resonances~\cite{Melde:2008yr}.

\section{CLASSIFICATION OF BARYON STATES}

The systematics found in the decay widths together with a detailed analysis of the
spin-flavor contents and spatial structures of the baryon wave functions allow for a
new and extended classification of the ground and resonant states into flavor
multiplets~\cite{Melde:2008yr}. The members of the flavor octets, decuplets, and singlets
following from our relativistic study are summarized in tables~\ref{tab:multiplet_oct}
to~\ref{tab:multiplet_singl}. In some instances we find assignments different from
the ones quoted by the PDG~\cite{PDBook}. A few resonances not considered by the PDG
reasonably fit into our multiplet classification (e.g., in the $\Sigma$ spectrum).
Of course, additional experimental evidences would be highly welcome, and are in fact
necessary, to confirm the assignments of certain states. For a thorough discussion of the
detailed properties of the members in each one of the multiplets, especially also with
regard to spatial probability density distributions in terms of Jacobi coordinates
between the constituent quarks, see ref.~\cite{Melde:2008yr}.

\renewcommand{\arraystretch}{1.3}
\begin{table}
\caption{Classification of flavor octet baryons. The denotation of the mass eigenstates is
made according to the nomenclature of baryon states seen in experiment. The superscripts denote the percentages of octet content as calculated with the GBE
RCQM~\cite{Glozman:1998ag}. States in bold face have either not been assigned by the
PDG~\cite{PDBook} or differ from their assignment.
\label{tab:multiplet_oct}
}
\vspace{0.2cm}
{\begin{tabular}{@{}l llll  @{}}
\hline
$(LS)J^P$& &&&\\
\hline
 $(0\frac{1}{2})\frac{1}{2}^+$
& $N(939)^{100}$  
& $\Lambda(1116)^{100}$
& $\Sigma(1193)^{100}$ &
$\Xi(1318)^{100}$
 \\
$(0\frac{1}{2})\frac{1}{2}^+$ 
& $N(1440)^{100}$  
& $\Lambda(1600)^{96}$
& $\Sigma(1660)^{100}$ 
& $\mbox{\boldmath$\Xi$}{\bf (1690)}^{100}$
\\
$(0\frac{1}{2})\frac{1}{2}^+$
& $N(1710)^{100}$  
&
& $\Sigma(1880)^{99}$ 
& 
 \\
$(1\frac{1}{2})\frac{1}{2}^-$
& $N(1535)^{100}$  
& $\Lambda(1670)^{72}$
& $\mbox{\boldmath$\Sigma$}{\bf (1560)}^{94}$
&  
 \\
$(1\frac{3}{2})\frac{1}{2}^-$
& $N(1650)^{100}$ 
& $\Lambda(1800)^{100}$ 
&$\mbox{\boldmath$\Sigma$}{\bf (1620)}^{100}$
& 
 \\
$(1\frac{1}{2})\frac{3}{2}^-$ 
& $N(1520)^{100}$
& $\Lambda(1690)^{72}$ 
& $\Sigma(1670)^{94}$ 
& $\Xi(1820)^{97}$
 \\
$(1\frac{3}{2})\frac{3}{2}^-$
& $N(1700)^{100}$ 
&
&$\mbox{\boldmath$\Sigma$}{\bf (1940)}^{100}$
& 
 \\
$(1\frac{3}{2})\frac{5}{2}^-$ 
&$N(1675)^{100}$  
& $\Lambda(1830)^{100}$
& $\Sigma(1775)^{100}$ 
& $\mbox{\boldmath$\Xi$}{\bf (1950)}^{100}$
 \\
\hline
\end{tabular}}
\end{table}

\renewcommand{\arraystretch}{1.3}
\begin{table}
\caption{
Classification of flavor decuplet baryons. Analogous notation as in
Table~\ref{tab:multiplet_oct}.
\label{tab:multiplet_decu}
}
\vspace{0.2cm}
{\begin{tabular}{@{}l llll @{}}
\hline
 $(LS)J^P$&&&&\\
\hline
$(0\frac{3}{2})\frac{3}2{}^+$
& $\Delta(1232)^{100}$ 
& $\Sigma(1385)^{100}$ 
& $\Xi(1530)^{100}$ 
& $\Omega(1672)^{100}$
 \\
$(0\frac{3}{2})\frac{3}{2}^+$
& $\Delta(1600)^{100}$ 
& $\mbox{\boldmath$\Sigma$}{\bf (1690)}^{99}$
& & 
  \\
$(1\frac{1}{2})\frac{1}{2}^-$
& $\Delta(1620)^{100}$  
&  $\mbox{\boldmath$\Sigma$}{\bf (1750)}^{94}$
& &
\\
$(1\frac{1}{2})\frac{3}{2}^-$
& $\Delta(1700)^{100}$ 
& & & 
 \\
\hline
\end{tabular}}
\end{table}

\renewcommand{\arraystretch}{1.3}
\begin{table}
\caption{Classification of flavor singlet baryons. Analogous notation as in
Table~\ref{tab:multiplet_oct}.
\label{tab:multiplet_singl}
}
\vspace{0.2cm}
{\begin{tabular}{@{}l llll @{}}
\hline
$(LS)J^P$ & & & & \\
\hline
$(1\frac{1}{2})\frac{1}{2}^-$
& $\Lambda(1405)^{71}$ & \hspace{49mm} & &
 \\
$(1\frac{1}{2})\frac{3}{2}^-$  
& $\Lambda(1520)^{71}$ & & &
 \\
$(0\frac{1}{2})\frac{1}{2}^+$ 
&  $\mbox{\boldmath$\Lambda$}{\bf (1810)}^{92}$ & & &
\\
\hline
\end{tabular}}
\end{table}

\section{Summary}

At present, RCQMs allow for a unified description of the light and strange baryon spectra
in reasonable agreement with experiment. This is especially true for the GBE RCQM, which
produces the right level orderings both in the $N$ and $\Lambda$ spectra, due to its
flavor-dependent hyperfine interaction. The description of the $\Lambda(1405)$, however, remains as a notorious problem.

The electroweak structure of the nucleons and other baryon ground states is well
described by covariant results following from the point-form approach; they are very
similar to the predictions by the II RCQM in the framework of the Bethe-Salpeter equation.

The strong decays cannot be explained by the PFSM results. Obviously considerable
refinements are necessary in the decay operator and/or resonance wave functions. Still,
the consistent pattern found in the results for the partial decay widths, in combination
with an analysis of the spin-flavor contents and spatial structures of the baryon
wave functions, gives useful hints for the classification of resonances into $SU(3)$
flavor multiplets.

\begin{theacknowledgments}
  This work was partially supported by the Austrian Science Fund (FWF-Project P19035).
  Some of the results discussed here have been achieved in previous collaborations with
  K. Berger, S. Boffi, L. Canton, L.Ya. Glozman, W. Klink, M. Radici, B. Sengl, K. Varga,
  and R.F. Wagenbrunn.
\end{theacknowledgments}

\bibliographystyle{aipprocl} 

\end{document}